\documentclass[doublecol]{epl2}
\usepackage{amsmath,amssymb}
\usepackage{graphicx}

\newcommand{\dd}{\mathrm{d}}

\newcommand{\pd}[2]{\frac{\partial #1}{\partial #2}}
\newcommand{\fd}[2]{\frac{\delta #1}{\delta #2}}
\newcommand{\mean}[1]{\langle #1 \rangle}

\newcommand{\IInt}[3]{\int_{#2}^{#3}\dd #1\;}

\renewcommand{\vec}[1]{\mathbf #1}

\newcommand{\al}{\alpha}

\newcommand{\gam}{\gamma}

\newcommand{\sig}{\sigma}

\newcommand{\F}{\mathcal F}
\newcommand{\dsm}{\dot s_\mathrm{med}}
\newcommand{\dst}{\dot s_\mathrm{tot}}

\newcommand{\eq}{^\mathrm{eq}}
\newcommand{\Ba}{B^{(\mathrm{a})}}
\newcommand{\Be}{B^{(\mathrm{e})}}
\newcommand{\Bp}{B^{(\mathrm{p})}}

\def\pjp{p_{n_j^+}}
\def\pjm{p_{n_j^-}}
\def\wpm{w_{n_j^+n_j^-}}
\def\wmp{w_{n_j^-n_j^+}}

\begin{document}

\title{Fluctuation-Dissipation Theorem in Nonequilibrium Steady States}

\author{Udo Seifert\inst{1} \and Thomas Speck\inst{2,3}}

\institute{                    
  \inst{1} {II.} Institut f\"ur Theoretische Physik, Universit\"at Stuttgart,
  70550 Stuttgart, Germany \\
  \inst{2} Department of Chemistry, University of California, Berkeley,
  California 94720, USA \\
  \inst{3} Chemical Sciences Division, Lawrence Berkeley National Laboratory,
  Berkeley, California 94720, USA
}
%\pacs{nn.mm.xx}{Second pacs description}
%\pacs{82.70.Dd}{Colloids}

\abstract{In equilibrium, the fluctuation-dissipation theorem (FDT) expresses
  the response of an observable to a small perturbation by a correlation
  function of this variable with another one that is conjugate to the
  perturbation with respect to \emph{energy}. For a nonequilibrium steady
  state (NESS), the corresponding FDT is shown to involve in the correlation
  function a variable that is conjugate with respect to \emph{entropy}. By
  splitting up entropy production into one of the system and one of the
  medium, it is shown that for systems with a genuine equilibrium state the
  FDT of the NESS differs from its equilibrium form by an additive term
  involving \emph{total} entropy production. A related variant of the FDT not
  requiring explicit knowledge of the stationary state is particularly useful
  for coupled Langevin systems. The \emph{a priori} surprising freedom
  apparently involved in different forms of the FDT in a NESS is clarified.}

\maketitle

% ---- Introduction ----

\section{Introduction}

Stochastic thermodynamics provides a framework for describing small driven
systems embedded in a heat bath of still well-defined
temperature~\cite{seif08}. Its crucial ingredients are a formulation of the
first law~\cite{seki97} and the notion of a stochastic entropy~\cite{seif05a}
both valid along single fluctuating trajectories. Using these concepts several
exact relations for distribution functions for quantities like
work~\cite{jarz97,croo00} and entropy
production~\cite{evan93,gall95,kurc98,lebo99,seif05a} have been
derived. Experimental tests have been performed on a variety of different
systems. Prominent examples include colloidal particles manipulated by laser
traps~\cite{wang02,blic06,andr07}, biomolecules pulled by AFM's or optical
tweezers~\cite{liph02,coll05}, and single defects observed using fluorescence
techniques~\cite{tiet06}. Reviews of this very active field can be found in
Refs.~\cite{bust05,rito07,seif08}.

A particularly interesting class of states are nonequilibrium steady states
(NESS) characterized both by a time-independent distribution and, as a result
of the external driving, nonvanishing currents. If such a NESS is perturbed by
an additional small external force or field, one can ask whether the response
of an observable of this system can be expressed by a correlation function
involving this observable and a second one. For slightly perturbed equilibrium
systems, such a connection between response and equilibrium fluctuation is
given by the well-known fluctuation-dissipation theorem
(FDT)~\cite{kubo,marc08}. The appropriate correlation function involves the
observable whose response is sought for and another variable that is conjugate
to the perturbation with respect to \emph{energy}. The first purpose of this
letter is to show that previously derived somewhat formal looking FDTs for
general Markovian processes~\cite{agar72,hang82} acquire a particularly simple
and transparent form using the concepts of stochastic thermodynamics: In a
nonequilibrium steady state, the response of a system to an additional small
perturbation is given by a correlation function of this observable and another
one that is conjugate to the perturbation with respect to stochastic
\emph{entropy}. Moreover, by expressing entropy production in the system as
the difference between total entropy production and that in the surrounding
medium, we can show that for a large class of systems the FDT in a NESS can be
obtained from the corresponding equilibrium form of the FDT by subtracting a
term involving \emph{total} entropy production. The latter result rationalizes
and generalizes recent results for diffusive systems driven by an external
force~\cite{hara05,spec06} or shear flow~\cite{spec09}, see
Refs.~\cite{blic07,gome09} for first experimental tests of such extended FDTs.
Adapting a recently introduced alternative strategy for deriving an
FDT~\cite{baie09}, we discuss a variant not requiring explicit knowledge of
the typically unknown stationary distribution. This form will be particularly
useful in simulations of coupled Langevin systems. Finally, we clarify the
\emph{a priori} surprising apparent freedom involved in different forms of the
FDT in a NESS. For a broader overview of the FDT especially in systems with
glassy dynamics, we refer to the review~\cite{cris03}.

% ---- FDT ----

\section{Derivation of the FDT}

For a derivation of these results in a fairly general setting, we consider an
arbitrary set of states $\{n\}$. These states could \emph{inter alia} signify
discrete spatial variables for a set of driven interacting diffusive degrees
of freedom obtained by spatially discretizing Langevin equations. Likewise,
they could code the states of any (bio)chemical reaction network. A transition
from state $m$ to $n$ happens with a rate $w_{mn}(h)$, which depends on an
external parameter $h$. The probability $\psi_m(t)$ for finding the system in
state $m$ at time $t$ obeys the master equation
\begin{equation}
  \label{eq:master}
  \partial_t\psi_m(t) = \sum_n L_{mn}\psi_n(t)
\end{equation}
with generator
\begin{equation}
  \label{eq:2}
  L_{mn} \equiv w_{nm} - \delta_{mn}\sum_k w_{mk},
\end{equation}
where we suppress the $h$-dependence. The stochastic trajectory $n(t)$ is a
sequence of jumps at times $\tau_j$ from $n_j^-$ to $n_j^+$. An observable $A$
acquires a time-dependence through $A(t) =\sum_mA_m\delta_{mn(t)}$ along such
a trajectory with mean $\mean{A(t)}=\sum_mA_m\psi_m(t)$. A particularly
relevant observable is the stochastic entropy~\cite{seif05a} defined as
$s(t)\equiv-\ln\psi_{n(t)}(t)$, where we set the Boltzmann constant to one
throughout the paper. For any fixed $h$, we denote the stationary distribution
by $p_n$, which obeys $\sum_n L_{mn} p_n=0$. In such a NESS, the stochastic
entropy becomes the observable
\begin{equation}
  \label{eq:s:ness}
  s(t) = -\ln p_{n(t)}, \quad\text{i.e.,}\quad
  s_n = -\ln p_n.
\end{equation}

We are interested in the response of the system to a small perturbation
$h(t)$, where the system is initially prepared in the NESS corresponding to
$h=0$ with stationary distribution $p_n^0$. The resulting general FDT has been
discussed first by Agarwal more than 30 years ago~\cite{agar72}. For
completeness and later reference, we briefly repeat its derivation. The
generator is expanded in powers of $h$,
\begin{equation}
  \label{eq:gen:series}
  \vec L(t) = \vec L^0 + h(t)\vec L^1,
\end{equation}
where $(\vec L)_{mn}=L_{mn}$. Due to the perturbation the distribution becomes
time-dependent and is obtained through formally solving the master
equation~(\ref{eq:master}),
\begin{equation}
  \label{eq:6}
  \psi_m(t) = \sum_n 
  \left[\exp\left\{ \IInt{\tau}{-\infty}{t} \vec L(\tau) \right\}\right]_{mn}
  p_n^0.
\end{equation}
The exponential function is to be understood as time-ordered. The mean
response of an observable $A$ is given through
\begin{equation}
  \label{eq:response}
  R_A(t_2-t_1) \equiv \left.\fd{\mean{A(t_2)}}{h(t_1)}\right|_{h=0}
  =\mean{A(t_2)B(t_1)},
\end{equation}
where we have introduced the generic form of the FDT equating the response
with a correlation function involving a second observable $B$ for which we
will find different forms $B^{(i)}$ labeled by different superscripts.

Using the expansion of the generator Eq.~(\ref{eq:gen:series}), one
immediately finds
\begin{equation}
  R_A(t_2-t_1) = \sum_{mn} A_m 
  \left[e^{\vec L^0(t_2-t_1)}\vec L^1\right]_{mn} p_n^0.
\end{equation}
Introducing the observable
\begin{equation}
  \label{eq:B:1}
  \Ba_m \equiv \sum_n L_{mn}^1(p^0_n/p^0_m),
\end{equation}
we thus can relate the response function to the correlations of $A$ with $B$
in the unperturbed NESS,
\begin{equation}
  \label{eq:fdt:agar}
  R_A(t_2-t_1) = \mean{A(t_2)\Ba(t_1)}.
\end{equation}
The observable Eq.~\eqref{eq:B:1} can be cast in a more explicit form
involving probabilities and rates by writing
\begin{equation}
  L_{mn}^1 = w_{nm}\al_{nm} - \delta_{mn}\sum_k w_{mk}\al_{mk}
\end{equation}
with the relative change of the rates
\begin{equation}
  \al_{mn} \equiv \partial_h\ln w_{mn}.
\end{equation}
Hence, from Eq.~\eqref{eq:B:1} we obtain the 'Agarwal' form 
\begin{equation}
  \label{eq:agar}
  \Ba_m = \sum_n (p^0_n/p^0_m)w_{nm}\al_{nm} - \sum_n w_{mn}\al_{mn}
\end{equation}
of the observable appearing at the earlier time in the
FDT~\eqref{eq:fdt:agar}.

\section{Role of stochastic entropy}

To establish the connection between the generic FDT as expressed in
Eq.~(\ref{eq:response}) and the stochastic entropy, we consider two NESSs
differing by an infinitesimal time-independent $h$ with distributions $p_n^0$
and $p_n^0+hp_n^1$, respectively. Using Eq.~(\ref{eq:gen:series}) and equating
terms linear in $h$, the relation
\begin{equation}
  \label{eq:ll}
  \sum_n L_{mn}^0 p_n^1 = -\sum_n L_{mn}^1 p_n^0
\end{equation}
holds. With the identification
\begin{equation}
  \label{eq:be}
  \left.\partial_h{s_m}(h)\right|_{h=0} = - p_m^1/p_m^0
\end{equation}
following from Eq.~(\ref{eq:s:ness}), we can express the response as
\begin{align}
  R_A(t_2-t_1) &= -\sum_{mkn} A_m \left[e^{\vec L^0(t_2-t_1)}\right]_{mk}
  L_{kn}^0 p_n^1 \\
  &= \pd{}{t_1} \sum_{mn} A_m \left[e^{\vec L^0(t_2-t_1)}\right]_{mn}
  (p_n^1/p_n^0)p_n^0 \\
  \label{eq:fdt}
  &= \pd{}{t_1} \mean{A(t_2)[-\partial_h s(t_1)]}= \mean{A(t_2)\Be(t_1)}
\end{align}
with
\begin{equation}
  \Be \equiv -\partial_h \dot s,
\end{equation}
where $s(t)$ is to be understood as observable in the sense of
Eq.~\eqref{eq:s:ness}. Thus, the response is given by a correlation function
involving as second variable the one being conjugate to the perturbation with
respect to stochastic entropy production. Formally similar relations have
previously been derived for general stochastic processes~\cite{hang82} and
also for the response of chaotic dynamics to changing initial
conditions~\cite{falc95}. The FDT~\eqref{eq:fdt} contains as a special case an
expression derived recently~\cite{pros09} following another route based on the
Hatano-Sasa relation~\cite{hata01}. The advantage of the present discussion
using the concepts of stochastic thermodynamics arises from the transparent
physical identification of the conjugate variable as stochastic entropy, which
constitutes our first main result.

As a consistency check, we consider the case where the steady state is a
genuine equilibrium state for $h=0$. In fact, two types of such systems should
be distinguished. Class I systems exhibit even for any small constant non-zero
$h$ a genuine equilibrium state like any magnetic system in the presence of a
perturbing magnetic field. For such systems, the stationary distribution is
given by the Boltzmann-Gibbs distribution
\begin{equation}
  \label{eq:5}
  p\eq_n(h)=\exp\{-[E_n(h)-\F(h)]/T\},  
\end{equation}
where $E_n(h)$ is the internal energy, $T$ the temperature of the heat bath,
and $\F(h)\equiv -T\ln\sum_n\exp(-E_n/T)$ the $h$-dependent free energy of the
system. The stochastic entropy obeys $s_n(h)=-\ln
p\eq_n(h)=[E_n(h)-\F(h)]/T$. Along a single trajectory, $\F(h)$ is constant
and hence we have
$T\partial_h{s(h)}|_{h=0}=\partial_h{E(h)}|_{h=0}$. Inserted into
(\ref{eq:fdt}), the FDT acquires its well-known equilibrium form
\begin{equation}
  \label{eq:fdt:eq}
  T R_A(t_2-t_1) =   \pd{}{t_1}\mean{A(t_2) [-\partial_hE(h)]_{h=0}(t_1)}
\end{equation}
involving the observable conjugate to $h$ with respect to energy.

Class II systems are in equilibrium at $h=0$ but driven into a NESS even at
constant small $h$. The paradigmatic example is a perturbation through shear
flow for which there is no corresponding $E(h)$ for any $h\neq0$. For such
systems one still has the FDT in the form~(\ref{eq:fdt}) but also in the
Agarwal form with~(\ref{eq:fdt:agar}).

Returning to the general case of perturbing an arbitrary NESS, the main
advantage of the present formulation using stochastic entropy becomes apparent
when we split up the entropy production in the NESS into two
terms~\cite{seif05a}
\begin{equation}
  \label{eq:bal}
  \begin{split}
    \dot s(t) &= -\sum_j\delta(t-\tau_j)\ln{\pjp\over \pjm}\\
    &= -\sum_j\delta(t-\tau_j)\ln{\wmp\over\wpm}
    \\ &\phantom{=}
    + \sum_j\delta(t-\tau_j)\ln{\pjm\wmp\over \pjp\wpm} \\
    &\equiv -\dsm(t)+\dst(t).
  \end{split}
\end{equation}
The first term on the right hand side, $\dsm$, denotes the entropy production
rate in the medium which can in many cases be identified with the heat flow
(divided by $T$) into the surrounding aqueous solution. The second term is the
total entropy production rate. On average, the latter is positive and
integrated over a finite time obeys the detailed fluctuation
theorem~\cite{seif05a}. Pulling in the time-derivative in Eq.~\eqref{eq:fdt}
and using Eq.~(\ref{eq:bal}), the correlation part of the FDT becomes a
difference between two terms involving entropy production,
\begin{equation}
  \label{eq:fdt2}
  R_A(t_2-t_1) = \mean{A(t_2)[\partial_h\dsm](t_1)}
  - \mean{A(t_2)[\partial_h\dst](t_1)}.
\end{equation}
Note that, although the mean total entropy production is always non-negative,
both terms can have either sign; see, e.g., Fig. 3a in Ref.~\cite{blic07}.

We now show that for systems with a genuine equilibrium state the first term
in Eq.~(\ref{eq:fdt2}) corresponds to the equilibrium form of the FDT and
hence the second term is an additive correction induced by the nonequilibrium
conditions. Such an additive structure was found
previously~\cite{cris03,diez05} without, however, giving the correction term a
transparent physical meaning. In the special case of a driven Langevin
particle~\cite{spec06}, the correction term has been identified as a
correlation function involving the local mean velocity and has later been
rationalized by observing the process in the locally comoving
frame~\cite{chet08,chet09}. Generally, the observable occurring in the first
term in Eq.~(\ref{eq:fdt2}) can be written
\begin{align}
  \label{eq:dhsm}
  \partial_h \dsm(t) &= \sum_j\delta(t-\tau_j)
  \partial_h \ln{\wmp\over\wpm} \\
  &= \sum_j \delta(t-\tau_j)\left[\alpha_{n_j^-n_j^+}-\alpha_{n_j^+n_j-}\right].
\end{align}
Since this observable appears in the correlation function at the earlier time
$t_1$, we have to average over all trajectories reaching state $n(t)$, i.e., $
\sum_j\delta(t-\tau_j) \al_{n^-_jn^+_j}\mapsto\sum_k p_kw_{kn(t)}\al_{kn(t)}$
and similarly for the $\al_{n^+_jn^-_j} $ term. Using furthermore for an
equilibrium system the detailed balance relation $p_nw_{nm}=p_mw_{mn}$, this
pre-averaged expression becomes Eq.~(\ref{eq:agar}). Hence, for any system in
equilibrium, the FDT~(\ref{eq:fdt:eq}) can also be written in the form
\begin{equation}
  \label{eq:b1}
  R_A(t_2-t_1) = \mean{A(t_2)[\partial_h\dsm]_{h=0}(t_1)}.
\end{equation}
If for the same system a NESS generated by a field $h$ is additionally
perturbed by the same field $h$, one can thus keep the equilibrium two-point
observable (but now evaluated under nonequilibrium conditions) and subtract
the second term that involves the observable conjugate to total entropy
production. This recipe for deriving the FDT in a NESS constitutes our second
main result. It sharpens and proofs the hypothesis formulated
in~\cite{spec09}.

% ---- Alternative ----

\section{Path weight approach}

The main virtue of the insight provided by the identifications of the various
terms entering the FDT as discussed above is of a conceptual nature. From a
practical point of view, quite generally, the FDT allows to infer response
functions from correlation functions, which are obtained more easily
both in experiments and in simulations. Necessary for a practical
implementation, however, is an explicit knowledge of the variable entering the
correlation function in the FDT. Any of the variants discussed above requires
knowledge of the stationary distribution $p_n$, which, in general, is not
known explicitly. A form of the FDT not requiring such knowledge can indeed be
derived following a recent suggestion~\cite{baie09} exploited there with a
different focus. In the following we derive the explicit form of such an FDT
both for general master equations and for coupled Langevin equations.

Consider the mean
\begin{equation}
  \label{eq:1}
  \begin{split}
    \mean{A(t)} &= \sum_{n(t)} A(t)P[n(t);h(t)] \\
    &= \sum_{n(t)} A(t)\frac{P[n(t);h(t)]}{P_0[n(t)]}P_0[n(t)],
  \end{split}
\end{equation}
where $P[n(t);h(t)]\equiv\exp\{-S[n(t);h(t)]\}$ is the path weight in the
perturbed system, $P_0[n(t)]\equiv\exp\{-S_0[n(t)]\}$ is the path weight in
the unperturbed NESS, and the sum runs over all paths $n(t)$. Taking the
functional derivative of Eq.~(\ref{eq:1}) with respect to $h(t_1)$ in order to
calculate the response function~\eqref{eq:response}, the result can again be
cast into the generic form
\begin{equation}
  \label{eq:fdt4}
  R_A(t_2-t_1) = \mean{A(t_2)\Bp(t_1)}.
\end{equation}
The correlation function is measured in the unperturbed NESS and the conjugate
observable is now expressed through
\begin{equation}
  \label{eq:3}
  \Bp(t_1) = -\left.\fd{S[n(t);h(t)]}{h(t_1)}\right|_{h=0}.
\end{equation}

\subsection{Discrete state space}

For a dynamics with rates which become time-dependent through an external
perturbation, $w_{mn}(h(t))$, the weight of a path starting at time $t_0<t_1$
with some initial weight $p_0\equiv p_{n(0)}$ and, after $N_j$ jumps at
$\tau_j$, ending at time $t>t_2$ is given by
\begin{equation}
  P[n(t);h(t)] = p_0 \exp\left\{-\IInt{\tau}{0}{t} r_{n(\tau)} \right\}
  \prod_{j=1}^{N_j} w_{n_j^-n_j^+}(h(\tau_j)),
\end{equation}
where $r_n\equiv\sum_{m\neq n}w_{nm}$ is the exit rate out of state
$n$. Hence, Eq.~(\ref{eq:3}) becomes
\begin{equation}
  \label{eq:BP}
  \Bp(t) = -\sum_kw_{n(t)k}\alpha_{n(t)k} +
  \sum_j\delta(t-\tau_j)\alpha_{n_j^-n_j^+}.
\end{equation}
Due to its second term, this observable is sensitive to the jumps. As its main
advantage, it does not require knowledge of the $p_n$ in contrast to all forms
discussed above. Since $\Bp$ appears in a correlation function at an earlier
time, the average in Eq.~(\ref{eq:fdt4}) over the jump times and the states
before the jumps can be performed as explained for Eq.~(\ref{eq:b1}) leading
to the form (\ref{eq:agar}), where the $p_n$ show up again explicitly. Whether
in a simulation or an experiment one uses the conjugate variable in the form
(\ref{eq:agar}) or in the form (\ref{eq:BP}) is a matter of convenience. The
second form is particularly suitable if one knows the rates and their
$h$-dependence but not the stationary distribution. If one knows (or can
measure) the latter more easily without knowing the rates explicitly, the form
(\ref{eq:fdt}) may even be more appropriate.

\subsection{Langevin dynamics}

The same approach can easily be followed for a set of $N$ coupled degrees of
freedom $\vec x\equiv(x_1,\dots,x_N)$ obeying a Langevin dynamics
\begin{equation}
  \label{eq:lang}
  \dot x_\alpha = \mu_{\alpha\beta} F_\beta + u_\alpha+ \zeta_\alpha  
\end{equation}
with correlations
\begin{equation}
  \mean{\zeta_\alpha(t_1) \zeta_\beta(t_2)} = 
  2T \mu_{\alpha\beta} \delta (t_2-t_1).  
\end{equation}
Here, $F_\beta(\vec x,h)$ is the force acting on the $\beta^{th}$ particle and
$\mu_{\alpha\beta}$ are mobilities connecting these degrees of
freedom. Advection by a fluid is included by a local velocity $u_\alpha(\vec
x,h)$. Here, and in the following we sum implicitly from 1 to $N$ over all
greek indices occurring twice.

The path weight becomes
\begin{equation}
  P[\vec x(t);h(t)] = \mathcal N \exp\left\{ -\frac{1}{4T} \IInt{t}{t_1}{t_2}
    \zeta_\al \mu^{-1}_{\al\beta} \zeta_\beta \right\}
\end{equation}
with $\zeta_\al$ replaced by the Langevin equation~(\ref{eq:lang}), where
$\cal N$ is an irrelevant normalization. Following the same steps as in the
discrete case, one finds for the conjugate observable\footnote{This result is
  independent of the chosen stochastic calculus (It\^o or Stratonovich) as
  long as the mobility coefficients are independent of $h$ and $h$ is a
  spatially homogeneous perturbation.}
\begin{equation}
  \label{eq:BP:lang}
  T\Bp = \frac{1}{2}(\dot x_\al-\mu_{\al\beta}F_\beta-u_\al)(\partial_h
  F_\al+\mu^{-1}_{\al\gam}\partial_h u_\gam).
\end{equation}
By using this variable in a simulation, one could thus predict the response
function for any perturbation by just calculating the corresponding correlation
function which constitutes our third main result.

% ---- Discussion ----

\section{Classification}

Different approaches allowed us to derive apparently different forms of the
FDT in a NESS. However, as the response function depends only on the
observable $A$, all these different correlation functions must be the
same. Hence, there is a class of equivalent observables $\{B\}$ leading to the
same value of the correlation function. In particular,
\begin{equation}
  \label{eq:equiv}
  \Ba \cong \Bp \cong \Be,
\end{equation}
where $\cong$ denotes the equivalence of these observables if they appear as
observable at the earlier time in a two-time correlation function taken in the
NESS. Moreover, in principle there are infinitely many variants of the FDT
since with $B^{(1)}\cong B^{(2)}$ any normalized linear combination $ [c_1
B^{(1)}+c_2B^{(2)}]/(c_1+c_2)$, with $c_{1,2}$ real, will be admissible.

The three main variants for the second variable $B$ appearing in the FDT can
more formally be classified as follows: (i) The Agarwal form
$\Ba$~(\ref{eq:agar}) is distinguished by the fact that it involves only state
variables and no time-derivative, i.e., no observables evaluated at jumps for
the discrete case or no velocity observables for Langevin systems. (ii) The
form $\Be$~(\ref{eq:be}) is the unique one where the $B$ observable is written
as a time-derivative of such a state variable, which turns out to be the
$h$-derivative of the stochastic entropy. (iii) The observable $\Bp$ is the
unique one not requiring explicit knowledge of the stationary distribution.

\section{Examples}

\begin{figure}[t]
  \centering
  \includegraphics[width=\linewidth]{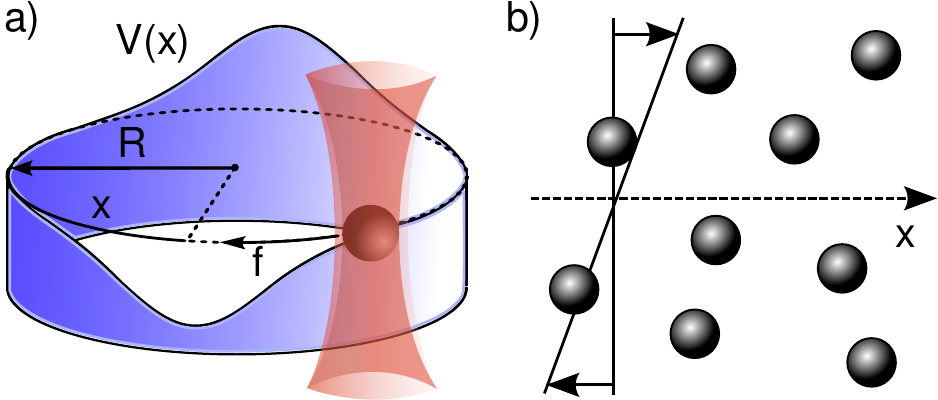}
  \caption{Example systems: a) Single driven colloidal particle and b) sheared
    colloidal suspension.}
  \label{fig:sys}
\end{figure}

We now illustrate both the general recipe for deriving an FDT for a NESS and
the equivalence of the main three forms of such an FDT for two previously
studied cases of driven Langevin dynamics sketched in Fig.~\ref{fig:sys}.

\subsection{Driven colloidal particle}

An overdamped colloidal particle with mobility $\mu_0$ is driven by a force
$f$ along a periodic potential $V(x)$ and hence subject to the total force
$F(x)=-V'(x)+f$ \cite{spec06}. The Langevin equation reads
\begin{equation}
  \dot x(t) = \mu_0 F(x) + \zeta(t)
\end{equation}
where the noise has zero mean and correlations
$\mean{\zeta(t_1)\zeta(t_2)}=2T\mu_0\delta(t_2-t_1)$. In the stationary state,
the constant probability current can be written as
\begin{equation}
\label{eq:j}
  j = \mu_0[F(x)p(x)-T\partial_x p(x)] \equiv \nu(x)p(x)
\end{equation}
with the local mean velocity $\nu(x)$. The three variants of the FDT can be derived as follows:

(i) A perturbation of the driving force $f$ corresponds to the operator
$L^1=-\mu_0 \partial_x$, which implies with the continuum version of
Eq. (\ref{eq:B:1}) and using Eq. (\ref{eq:j}) the Agarwal form
\begin{equation}
  \label{eq:7}
  T\Ba = \nu(x) - \mu_0 F(x)
\end{equation}
distinguished by the fact that it does not involve any fluctuating velocity
variable.

(ii) The stochastic entropy production rate in the medium is given
by~\cite{seif05a,spec08}
\begin{equation}
  \dsm = (1/T)[-V'(x)+f]\dot x.
\end{equation}
Then $T\partial_f\dsm=\dot x$, which corresponds to the equilibrium form of
the FDT since $x$ is the conjugate variable of the force with respect to
energy. In a former study~\cite{spec06}, we found that the excess correlates
$A$ with the local mean velocity $\nu(x)$ and hence
\begin{equation}
  \label{eq:ring2}
  T\Be = \dot x -\nu(x).
\end{equation}
Since the total entropy production rate obeys $\dst=\dot x\nu(x)/(\mu_0T)$, we
get the by no means obvious equivalence
\begin{equation}
  T\partial_f\dst = (\dot x/\mu_0)\partial_f\nu(x) \cong \nu(x)
\end{equation}
when appearing in a correlation function at earlier times.

(iii) Finally, specializing the observable~(\ref{eq:BP:lang}) to this case, we
get as conjugate observable
\begin{equation}
  \label{eq:8}
  T\Bp = \frac{1}{2}(\dot x-\mu_0F).
\end{equation}
This third form, which does not require knowledge of the stationary
distribution, can easily be expressed as a linear combination of the first two
variants.

\subsection{Sheared suspension}

For a second illustration, consider $N$ colloidal particles with potential
energy $U(\{\vec r_k\})$ composed of pairwise interactions immersed into a
fluid which is sheared \cite{spec09}. We neglect hydrodynamic interactions
through $\mu_{kl}=\mu_0\delta_{kl}$. The velocity profile of the fluid is
assumed to be $\vec u(\vec r)=\gam(y,0,0)^T$, where $\gam$ is the strain rate
and $\vec r=(x,y,z)^T$. Response relations in sheared suspensions have
recently been studied also in the framework of mode-coupling
theory~\cite{krug09}.  In our formalism, the three main variants of the FDT
specialize to the following expressions:

(i) The 'Agarwal' form when perturbing the system through a small variation of
the strain rate reads
\begin{equation}
\label{eq:shear1}
  T\Ba = -T\sum_k y_k \pd{}{x_k}\ln p = \sig_{xy} - \bar\sig_{xy},
\end{equation}
where we have introduced the microscopic stress due to particle interactions
\begin{equation}
  \sig_{xy} \equiv \sum_k y_k \pd{U}{x_k}.
\end{equation}
The position of the $k$-th particle is $\vec r_k$ and the sum runs over all
particles. The second term is
\begin{equation}
  \bar\sig_{xy} \equiv \sum_k y_k\pd{}{x_k}[U+T\ln p]
  = -\frac{1}{\mu_0}\sum_k y_k(\nu_{k,x}-\gam y_k),
\end{equation}
where $\boldsymbol\nu_k$ is the local mean velocity of the $k$-th particle.

(ii) The medium entropy production rate is~\cite{spec08}
\begin{equation}
  \dsm = (1/T) \sum_k[\dot{\vec r}_k-\vec u({\vec r}_k)]\cdot[-\nabla_k U]
\end{equation}
and hence
\begin{equation}
 T\partial_\gam\dsm = \sum_k y_k \pd{U}{x_k} = \sig_{xy}
\end{equation}
becomes the microscopic stress. Comparing with the form~(\ref{eq:fdt2}) we can
therefore deduce that $T\partial_\gam\dst\cong\bar\sig_{xy}$. Here, the
Agarwal form and the one based on entropy production become identical because
$\sig_{xy}$ depends only on $\{\vec r_k\}$ and not on velocities. As noted
earlier, $\sig_{xy}$ cannot be written as a conjugate observable with respect
to energy since for any $\gam\neq0$ the system reaches a genuine NESS instead
of another equilibrium state.

(iii) Following the path weight approach, we obtain by specializing
Eq.~(\ref{eq:BP:lang})
\begin{equation}
  \label{eq:shear2}
  \begin{split}
    T\Bp &= \sum_k \frac{1}{2\mu_0}\left(\dot x_k+\mu_0\pd{U}{x_k}
      -\gam y_k\right)y_k \\
    &= \frac{1}{2}\sig_{xy} + \frac{1}{2\mu_0} \sum_k y_k(\dot x_k-\gam y_k),
  \end{split}
\end{equation}
which can easily be used in simulations.

A straightforward combination of the forms~(\ref{eq:shear1})
and~(\ref{eq:shear2}) leads to
\begin{equation}
  T[2\Bp-\Ba] = \frac{1}{\mu_0}\sum_k y_k(\dot x_k-\nu_{k,x})
\end{equation}
as another admissible observable appearing in the FDT for a perturbation of
the strain rate $\gam$. This form stresses the relevance of deviations from
the local mean velocity in analogy to the form (\ref{eq:ring2}) for the driven
colloidal particle.

% ---- Conclusions ----

\section{Summary}

We have discussed the FDT for Markov processes driven into a NESS. The
response of any observable to a perturbation can be expressed by a correlation
function involving the observable conjugate to the perturbation with respect
to stochastic entropy production. By expressing the later through entropy
production in the medium and the total one, the nonequilibrium form of the FDT
can be written as the equilibrium one and an additive correction.
Alternatively, we have derived a variant of the FDT not requiring explicit
knowledge of the stationary distribution. A classification based on the
concept of equivalent observables shows that, quite generally, there exist two
linearly independent forms of the FDT from which one can derive three main
variants distinguished by the nature of the conjugate observable. Our general
framework not only rationalizes previous results obtained for case studies but
should also be helpful in studying the FDT in specific future applications
like, e.g., driven biochemical networks or active biophysical systems.

% ---- Acknowledgements ----

\acknowledgments

U.S. acknowledges funding through DFG project SE1119/3-1 and ESF network
EPSD. T.S. acknowledges funding through Alexander von Humboldt foundation and
the Helios Solar Energy Research Center which is supported by the Director,
Office of Science, Office of Basic Energy Sciences of the U.S. Department of
Energy under Contract No.~DE-AC02-05CH11231. We are grateful to J. Mehl for
providing us with figure~\ref{fig:sys}a).

% ---- Bibliography ----

\end{document}